\documentclass[aps,preprint,showpacs,preprintnumbers,amsmath,amssymb]{revtex4}

\usepackage{graphicx}
\begin{document}
\title{
 Self-consistent fragmented excited states of trapped condensates.}
\author{L.S.\ Cederbaum} 
\author{A.I.\ Streltsov}
\affiliation{Theoretische Chemie,Universit\"at Heidelberg, D-69120 Heidelberg, Germany}
\date{\today}
\begin{abstract}
Self-consistent excited states of condensates are solutions of the Gross-Pitaevskii (GP) equation 
and have been amply discussed in the literature and related to experiments.
By introducing a more general mean-field which includes the GP one as a special case,
we find a new class of self-consistent excited states. In these states
macroscopic numbers of bosons reside in different one-particle functions, i.e., the states are fragmented.
Still, a single chemical potential is associated with the condensate.
A numerical example is presented, illustrating that the energies of the
new, fragmented, states are much lower than those of the GP excited states, and that they 
are stable to variations of the particle number and shape of the trap potential.
\end{abstract}
\pacs{03.75.Hh,03.65.Ge,03.75.Nt}
\maketitle

Numerous properties of dilute Bose-Einstein condensates are well discussed 
by the Gross-Pitaevskii (GP) equation \cite{1,2}. For reviews see, e.g. \cite{3,4,5}. 
If the condensate is trapped, this equation, which is equivalent to the so called nonlinear Schr\"odinger equation,
possesses a discrete spectrum of stationary states. Being a nonlinear equation which has to be solved 
self-consistently, the solutions of the GP equation are generally called self-consistent states.
The properties of GP self-consistent {\it excited} states and their formation 
have been amply discussed, see, e.g. \cite{6,7,8,9,10,11,12,13,14}. The GP equation supports many well-known self-consistent 
excitations, such as vortex states \cite{15,16,17,18,19}, and bright and dark solitons \cite{20,21,22,23,24}.
The observation of solitons in trapped Bose gases \cite{25,26} provides a striking manifestation
of nonlinear atom optics \cite{27}.

By introducing a more general mean-field which includes the GP one as a special case, we find another
class of self-consistent excited states. Numerical examples show that the corresponding energies
are substantially lower than those of the GP excited states.
To avoid misunderstanding we mention that self-consistent excited states are physically distinct from collective 
(or particle-hole) excitations. The latter correspond to small oscillations around a given state and are described
by mean-field {\it linear}-response theories based on the Bogoliubov approximation \cite{28,29}.

We consider a system of N identical bosons. The corresponding GP equation is the mean-field equation
of this system with a $\delta$-function interaction potential 
$ W(\vec{r}_i-\vec{r}_j)=\lambda_0\,\delta(\vec{r}_i-\vec{r}_j) $, where $\vec{r}_i$ is the position of the i-th
boson and the nonlinear parameter $\lambda_0$ is related to the s-wave scattering length of the bosons \cite{4}. 
The GP ansatz further assumes the wave function $\Psi_{GP}$ to be a product of identical spatial one-particle 
functions (orbitals) $\varphi$:
$\Psi_{GP}(\vec{r}_1,\vec{r}_2,\ldots,\vec{r}_N)= \varphi(\vec{r}_1)\varphi(\vec{r}_2)\cdots\varphi(\vec{r}_N)$.
With this wavefunction the total energy $E_{GP}=<\Psi_{GP}|\hat{H}|\Psi_{GP}>$ is obtained as the
expectation value of our Hamiltonian $\hat{H}$ and reads
\begin{equation}
E_{GP}=N\{\int\varphi^* h\,\varphi\,d\vec{r}+\frac{\lambda_0(N-1)}{2}\int|\varphi|^4\,d\vec{r}\},
\end{equation}
where $h(\vec{r})=\hat{T}\,+\,\hat{V}(\vec{r}) $ is the unperturbed one-particle Hamiltonian consisting of the
trap potential $\hat{V}(\vec{r})$ and the kinetic energy operator $\hat{T}$.
By minimizing the energy (1) one obtains the GP equation:
\begin{equation}
\{\:h(\vec{r})+\lambda_0(N-1)|\varphi(\vec{r})|^2\}\,\varphi(\vec{r})=\,\mu_{GP}\,\varphi(\vec{r}).  
\end{equation}
Each self-consistent solution of this eigenvalue equation determines $\varphi$ and the chemical potential $\mu_{GP}$.
The ground state of the system, of course, corresponds to the solution with the lowest energy $E_{GP}$.
The other solutions describe the self-consistent excited states of the system.

In the absence of the mutual interaction  between the bosons, all bosons reside in a single spatial orbital in the 
ground state of the system. The GP ansatz for the wavefunction $\Psi_{GP}$ is thus very appealing
for the ground state and has indeed been very successful in explaining many observations.
We have reason to assume, however, that not all relevant macroscopic excited states are describable by $\Psi_{GP}$.
The general  wavefunctions $\Psi$ of N {\it non}-interacting bosons is a product of orbitals which can all be different.
Since the bosons are identical, this product must be symmetrized and several bosons may reside in the same spatial orbital.
We may put $n_1$ bosons in orbital $\phi_1$, $n_2$ in orbital $\phi_2$ and so on.
For transparency of presentation we restrict ourselves in the following to two orbitals $\phi_1$ and $\phi_2$
with particle occupations $n_1$ and $n_2$ and $n_1+n_2=N$. The extension to more orbitals is straightforward. 
The wavefunction now reads
\begin{eqnarray}
\Psi(\vec{r}_1,\ldots,\vec{r}_N)= 
\hat{\cal S}\phi_1(\vec{r}_1)\cdots\phi_1(\vec{r}_{n_1})\phi_2(\vec{r}_{n_1+1})
\cdots\phi_2(\vec{r}_{n_1+n_2}),
\end{eqnarray}
where $\hat{\cal S} $ is the symmetrizing operator.

By definition the most general mean-field energy is $E=<\Psi|\hat{H}|\Psi>$ where $\Psi$ is the above 
discussed wavefunction. It is
evaluated to give \cite{30}
\begin{eqnarray}
 E = n_1 h_{11} + n_2 h_{22} + \lambda_0 \frac{n_1(n_1-1)}{2}\int|\phi_1|^4 d\vec{r}+
 \nonumber  \\  \lambda_0 \frac{n_2(n_2-1)}{2}\int|\phi_2|^4 d\vec{r}+
2 \lambda_0 n_1 n_2 \int|\phi_1|^2 |\phi_2|^2 d\vec{r},
\end{eqnarray}
where $h_{ii}=\int\,\phi_i^*\,h\,\phi_i\,d\vec{r}$ is the usual one-particle energy.
We now minimize this energy with respect to the orbitals under the constraints 
that they are orthogonal and normalized, i.e.,
\begin{eqnarray}
 <\phi_1|\phi_2>=0, <\phi_1|\phi_1>=<\phi_2|\phi_2>=1.
\end{eqnarray}
This leads to the following set of coupled equations for the optimal orbitals:
\begin{eqnarray}
\{\:h(\vec{r})+\lambda_0 (n_1-1)|\phi_1(\vec{r})|^2+2 \lambda_0 n_2|\phi_2(\vec{r})|^2 \}\,
\phi_1(\vec{r}) =  \nonumber & & \\ 
= \mu_{11}\,\phi_1(\vec{r})+\mu_{12}\,\phi_2(\vec{r}) \nonumber & & \\
\{\:h(\vec{r})+\lambda_0 (n_2-1)|\phi_2(\vec{r})|^2+2 \lambda_0 n_1|\phi_1(\vec{r})|^2 \}\,
\phi_2(\vec{r}) = \nonumber & & \\
= \mu_{22}\,\phi_2(\vec{r})+\mu_{21}\,\phi_1(\vec{r}).& & 
\end{eqnarray}
These equations should not be confused with those for condensates made of two types of bosons.
The $\mu_{ij}$ are the Lagrange parameters due to the above mentioned constraints;
$\mu_{11}$ and  $\mu_{22}$ are related to the normalization constraints and the introduction of $\mu_{12}$ and $\mu_{21}$
enforces orthogonality of $\phi_1$ and $\phi_2$. We mention that $n_1\mu_{12}=n_2\mu_{21}$.
In general, $\mu_{12} \neq 0 $ unless $\phi_1$ and $\phi_2$ 
are of different spatial symmetry.

By setting $n_2=0$ we readily see that the general mean-field wavefunction $\Psi$ in (3)
contains the GP wavefunctions $\Psi_{GP}$ as a special case. The energy $E$ in (4) then reduces 
to the expression for $E_{GP}$ in (1). Because of $n_1\mu_{12}=n_2\mu_{21}$, the parameter $\mu_{12}$ vanishes and the 
first equation in  (6) is then nothing but the GP equation (2).
Contrary to the GP equation (2), the general mean-field equations (6) are - owing to the presence 
of $\mu_{12}$ and $\mu_{21}$ - {\it not} eigenvalue equations. This is a generic property of the latter equations.
Unless either one of the particle occupations vanishes, or $n_1=n_2=N/2$, 
or $\phi_1$ and $\phi_2$ have different symmetry, or
$\mu_{11}=\mu_{22}$, the off-diagonal Lagrange parameters $\mu_{12}$ and $\mu_{21}$ cannot be removed from (6) by 
linear transformations. 

While we cannot attribute a physical observable to the off-diagonal Lagrange parameters $\mu_{12}$ and $\mu_{21}$
we can do so for the diagonal ones. As can be seen from (1), the energy needed to remove a boson from the condensate
without changing the orbital $\varphi$ is, within GP mean-field, given by $E_{GP}(N)-E_{GP}(N-1)=\mu_{GP}$.
Analogously, we can compute within the present general mean-field the energy needed to remove a boson from
orbital $\phi_1$ and that from orbital $\phi_2$. Recalling that $\phi_1$ and $\phi_2$ are orthogonal, we readily find
from (4) and (6) the identities $\mu_{11}=E(n_1,n_2)-E(n_1-1,n_2)$ and $\mu_{22}=E(n_1,n_2)- E(n_1,n_2-1)$.
Clearly, $\mu_{11}$ and $\mu_{22}$ can be viewed as chemical potentials of the $\{\phi_1\}$- and  $\{\phi_2\}$-boson
manifolds, respectively.
 
Next, we can make use of the fact that the energy $E$ and the 
orbitals $\phi_1$ and $\phi_2$ depend on the particle occupation
$n_1$ and treat this occupation as a variational parameter, i.e., search for its optimal value which
makes the energy stationary. Interestingly, one can show for macroscopic occupancies $n_1,n_2\gg1$ that
at any extremum of the energy as a function of $n_1$ the two quantities $\mu_{11}$ and $\mu_{22}$ coincide \cite{30}:
\begin{eqnarray}
\mu_{11}=\mu_{22}.
\end{eqnarray}
This is a relevant finding for understanding the concept of a condensate and the meaning of its self-consistent states.
The system described by the wavefunction (3) consists of two (or more) subsystems each possessing its own chemical
potential. This may contradict the picture one usually has of a condensate. It is thus important to note
that these, in general different, chemical potentials become identical at the optimal occupations restoring thereby
the picture of a condensate. With this in mind, we define as self-consistent condensate states of the present
mean-field only those solutions of (6) at which the chemical potentials $\mu_{11}$ and $\mu_{22}$ are identical.
These correspond to the extrema of the energy $E$ as a function of the boson occupation $n_1$.

Let us refer here to the problem of fragmentation often discussed in the literature, see refs. \cite{31,32} 
and references therein. A condensate is fragmented if its reduced one-body density matrix has two or more macroscopic
eigenvalues (the total number of particles N is assumed large). Until now fragmentation has not been found 
for trapped condensates. Our present ansatz has the potential to describe fragmentation on the mean-field level.
If the optimal boson occupations $n_1$ and $n_2$ are both macroscopic for large N, fragmentation in the respective
self-consistent state indeed takes place.
Consequently, we call such states which are beyond reach of the GP equation (2), 
{\it self-consistent fragmented states}.

We shall demonstrate in the following that the present mean-field supports 
self-consistent fragmented excited states. Moreover, these states can be at
{\it much lower energy than all} the GP self-consistent excited states.
As an example we choose a repulsive condensate $(\lambda_0>0)$ in an one-dimensional double-well potential.
Since equations (6) have very recently been derived, the numerical procedures to evaluate them are not yet sufficiently
developed to allow for computations in three dimensions. We have reason to believe, however, that similar results will
be obtained in more dimensions.

In our example the potential   
\begin{equation}
V(x)=c+\frac{\omega}{2}(x^2-\sqrt{a^2+(\Delta-2xx_0)^2})  
\end{equation}
describes two wells separated by a barrier. The depths of these wells differ by a bias $\approx\Delta$.
The parameter $c$ is chosen to set the bottom of the potential equal to zero. The coordinate $x$ is dimensionless and 
all energies and $\lambda_0$ are now in units of the frequency $\omega$.
The equations for the optimal orbitals have been evaluated numerically 
by using the DVR method \cite{33}. Starting from an initial guess for $\phi_1$ and $\phi_2$, for instance,
the solutions of the equations with slightly different boson occupations, the equations are iterated 
until self-consistency is achieved. To accelerate the convergence of the calculations we have
employed an energy-shift technique as often done in Hartree-Fock electronic structure computations \cite{34}.
The GP equation has been evaluated by solving the corresponding eigenvalue equation self-consistently using the DVR method.


Having obtained the optimal orbitals for the present and GP mean-fields, we have computed
the energies $E$ and $E_{GP}$. Computations have been carried out systematically for the ground and the first few
excited states using various values of the coupling constant $(\lambda=\lambda_0N)$ (note that $N\gg1$)
and of the parameters appearing in the potential $V(x)$ in (8). For each set of these values the optimal orbitals
and the mean-field energy $E$ have been determined as a function of the boson occupation $n_1$.
A typical example is shown in fig.1. 
In this figure the mean-field energy per particle $E/N$ is depicted as a
function of the fractional occupation $n_1/N$ for $\lambda=2.5$.
For comparison, the energies per particle $E_{GP}/N$ of the GP ground and first excited state are indicated
as well. Two disjoint $E/N$ curves are seen in the figure and we concentrate first on the one at lower energy.
Starting the calculations at $n_1=N$, the energies $E$ and $E_{GP}$, of course, coincide. We then continuously
lowered the value of $n_1/N$ and obtained a smooth $E/N$ curve.
This curve exhibits two maxima and a minimum which is, of course, located between them. Following (7) there is 
only one value for the chemical potential at each of these extrema.
Accordingly, the states corresponding to these extrema are self-consistent fragmented excited states
of the condensate.

The two maxima of $E(n_1)$ correspond to metastable states. Any variation of the boson occupation $n_1$
lowers the energy favoring a "decay" into either the GP ground state which is also the ground state of 
the present mean-field  or to the state corresponding to the minimum of $E(n_1)$.
The latter state is stable with respect to variations of $n_1$, or briefly a stable state.
A substantial energy barrier has to be overcome in order to lower the energy and to destabilize this state.
Using equations (4-6) we can show that 
\begin{equation}
\frac{d E}{d n_1}=\mu_{11}-\mu_{22}
\end{equation}
holds. This physically appealing relation adds insight into the understanding of the curves $E/N$
depicted in fig.1. For large N the quantity $\mu_{11}-\mu_{22}$ is the energy gain obtained by removing a
boson from the $\phi_1$-manifold and adding it to the $\phi_2$-manifold. $\mu_{11}-\mu_{22}$ is thus the 
"driving force" for the flow of bosons between the two boson subsystems. At the extrema of $E$ as a function of
the boson occupation one recovers (7).

The orbitals $\phi_1(x)$ and $\phi_2(x)$ for the stable fragmented excited state discussed above are depicted
in fig.2A. For comparison also the orbitals $\varphi$ for the ground and for the first excited GP state are shown.
The ground state GP orbital is as usual nodeless and consists of two humps related to the two wells of the
potential. The first excited state GP orbital exhibits a node close to the top of the barrier.
Each of the orbitals $\phi_1$ and $\phi_2$ of the fragmented stable excited state also possesses a node, but these
nodes are located at distinct different sites. $\phi_1$ is rather localized at the deeper well and its
node is close to the minimum of the other well. The reverse situation holds for $\phi_2$.
That these nodes are well separated can be understood by comparing the density per particles 
$\rho=( n_1|\phi_1|^2+ n_2|\phi_2|^2)/N$ with the respective quantities $\rho_{GP}= |\varphi|^2$ for the GP states.
These densities are shown in fig.2B together with the trap potential $V(x)$. Since the bosons in our example
repeal each other, the density would like to spread over space to reduce this repulsion. Indeed, in the ground state
the density is substantial in between the wells. Due to the node of the excited GP state, the density vanishes close to 
the top of the barrier, and as a consequence the density is enhanced inside the wells. This inevitably leads 
to a substantial increase of the energy $E_{GP}$ for the excited state. 
Fragmentation assigns different nodes to different orbitals and hence the density can penetrate 
the region of the barrier as seen in fig.2B. Consequently, the energy of the fragmented excited state is much
lower than that of the GP excited state as seen in fig.1.


It should be stressed that the above findings depend only weakly on the  shape of the double-well potential.
We have performed numerous calculations varying the coupling constant $\lambda$, the bias $\Delta$ 
and other parameters of the trap potential.
In particular, if we choose the potential to be symmetric (bias $\Delta=0$), the same conclusions as above can be drawn.
Furthermore, enlarging the coupling constant $\lambda=\lambda_0N$ results 
in even more prominent minimum and maxima of the energy curves as a function of $n_1/N$.

Let us briefly discuss the short $E/N$ curve in fig.1. We have obtained this curve by solving (6) starting from the
GP excited state orbital, i.e., from $n_2=N$, and increasing $n_1$ continuously.
As seen in the insert which shows this curve on an enlarged scale, the curve exhibits a maximum thus describing another
fragmented metastable state. It is also seen that the curve has a boundary minimum at $n_1=0$, 
indicating that the corresponding GP excited state is stable. The energy barrier involved is, however, very low.
We remark that at large values of $\lambda (\gtrsim8.0)$, this barrier disappears and we find a boundary maximum,
i.e., the GP state becomes metastable. The $E(n_1)$ curve then acquires a minimum at $n_1\neq0 $ giving rise
to a new stable fragmented state.

In conclusion, self-consistent fragmented excited states of condensates exist.
These states can be either metastable or stable with respect to the flow of bosons from one fragment 
(manifold of bosons residing in one orbital) to another one. 
In the class of examples studied here energies of fragmented states are much lower 
than those of the excited GP states. Due to the transparent physics behind populating several orbitals in
excited states, we expect such states to exist also in two and three dimensions and in other trap potentials.

The authors acknowledge useful discussions with Ofir Alon.


\pagebreak
\begin{figure}
\includegraphics[width=13.2cm]{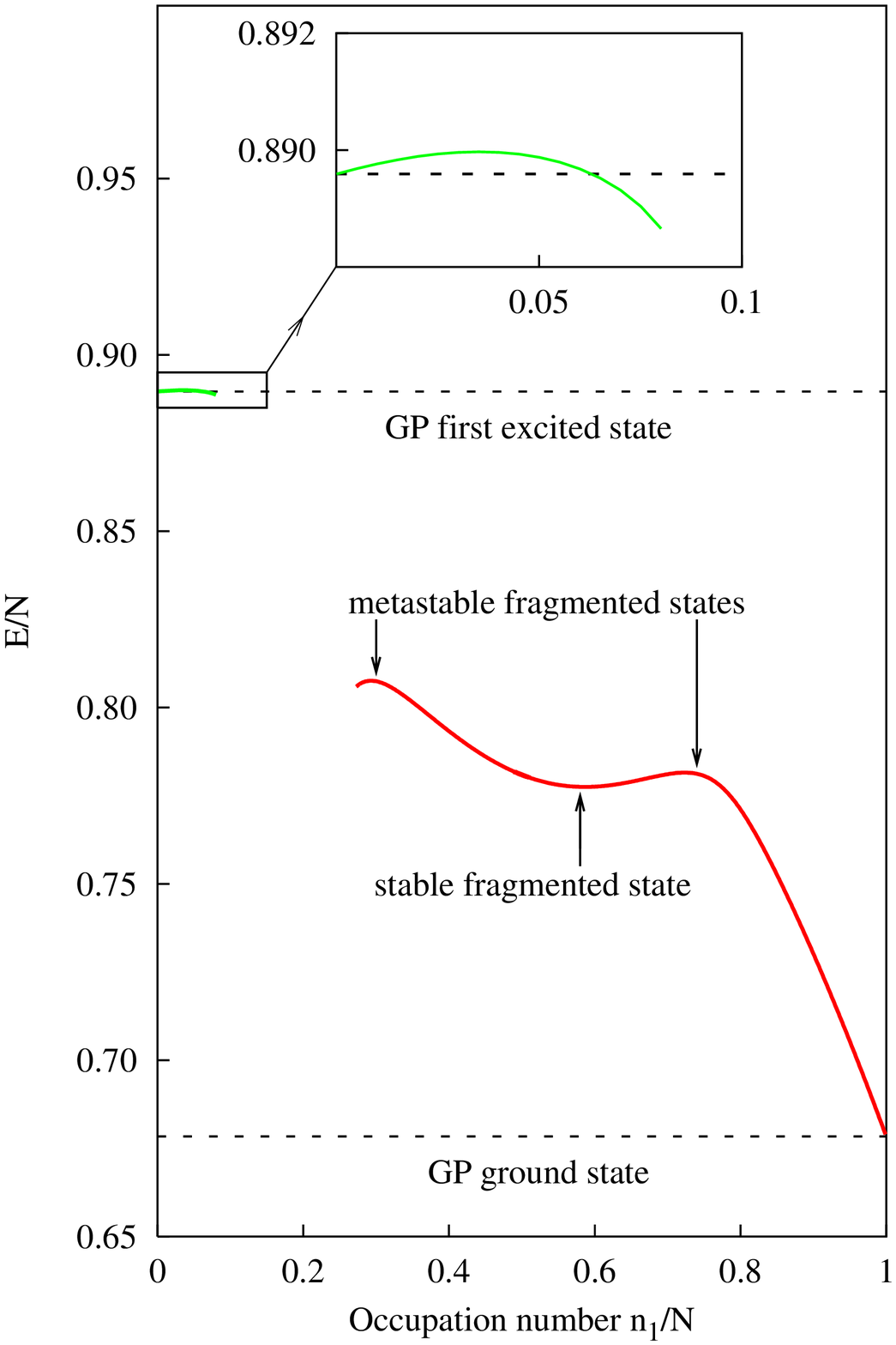}
\caption{
Energy per particle $E/N$ as a function of the fractional
boson occupation $n_1/N$ for the coupling constant $\lambda=\lambda_0\,N=2.5$.
For comparison, the energies $E_{GP}/N$ of the GP ground and first excited state are indicated. Two curves are shown,
starting at the respective GP energies. The following parameters of the trap potential (8)
have been used: $\Delta=0.1$, $a=0.04$  and $x_0=1.5$.}
\label{fig1}
\end{figure}

\begin{figure}
\includegraphics[width=12.2cm]{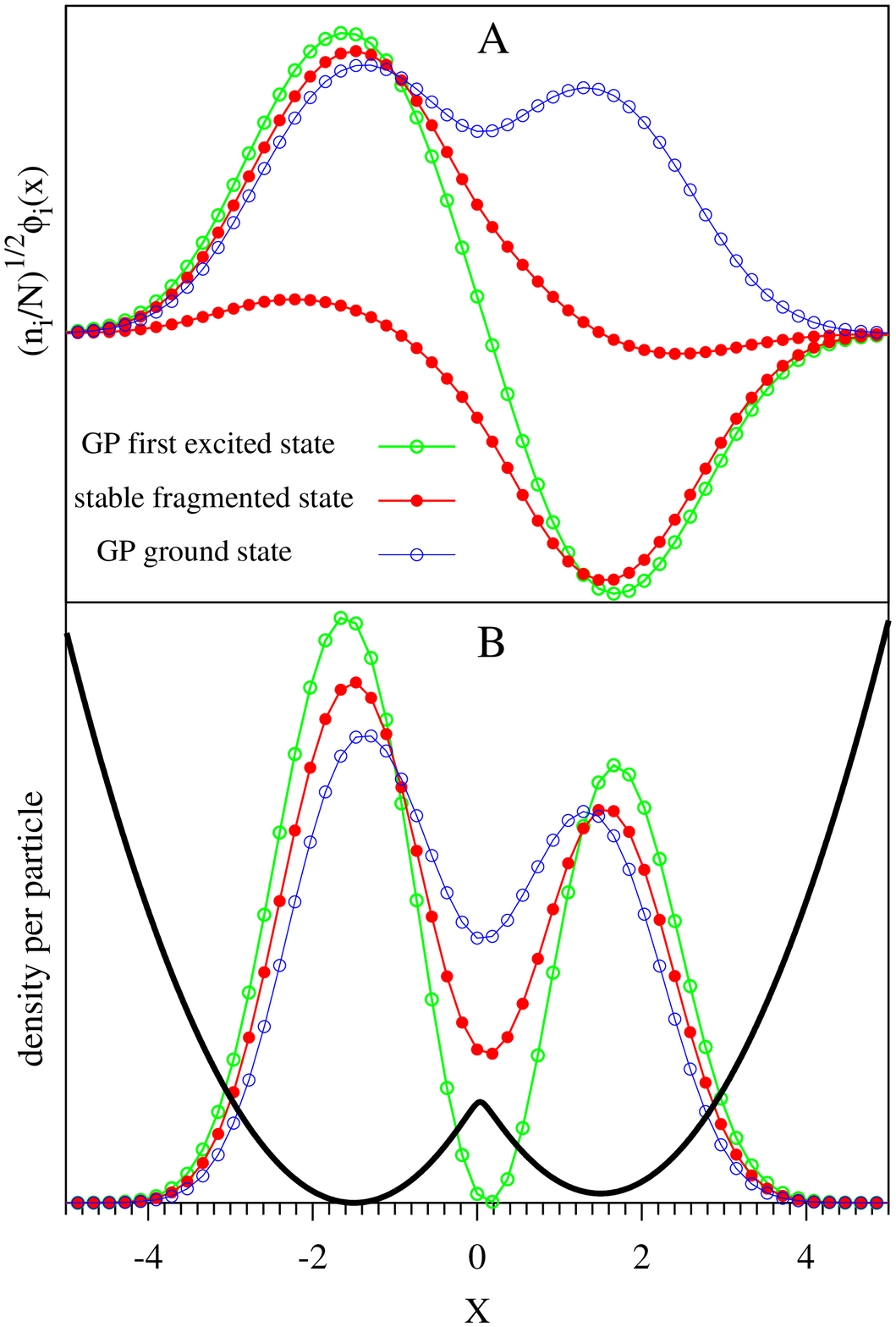}
\caption{
{\bf A:} The orbitals $\phi_1(x)$ and $\phi_2(x)$ corresponding to the stable fragmented excited state of fig.1.
(for convenience  $(n_i/N)^{1/2}\phi_i(x)$ are shown) in comparison with the orbitals $\varphi$
corresponding to the GP ground and first excited state.\\
{\bf B:} The densities per particle $\rho=( n_1|\phi_1|^2+ n_2|\phi_2|^2)/N$ of the stable fragmented excited state
and $\rho_{GP}= |\varphi|^2$ of the GP ground and first excited state. Also shown is the trap potential $V(x)$
(for parameters, see caption of fig.1). The values of the potentials have been scaled by 1/20.}
\label{fig2}
\end{figure}

\end{document}